\documentclass{ws-ijmpd}
\usepackage[super,compress]{cite}
\begin{document}

\markboth{Authors' Names}
{Black hole remnants due to Planck-length deformed QFT}


%
\catchline{}{}{}{}{}
%

\title{Black hole remnants due to Planck-length deformed QFT}

\author{Alain~R.~P.~Dirkes}
\address{Frankfurt Institute for Advanced Studies (FIAS), Johann Wolfgang 
Goethe Universit\"at, Ruth-Moufang-Strasse 1, Frankfurt am Main, D-60438, 
Germany \\ dirkes@fias.uni-frankfurt.de}

\author{Michael~Maziashvili}
\address{School of Natural Sciences and Engineering \\ Ilia State University, 
3/5 Cholokashvili Ave., Tbilisi 0162, 
Georgia \\ maziashvili@iliauni.edu.ge}

\author{Zurab~K.~Silagadze}
\address{Budker Institute of Nuclear Physics SB RAS and \\ 
Novosibirsk State University,
630 090, Novosibirsk, Russia \\ silagadze@inp.nsk.su}

\maketitle

\begin{history}
\received{Day Month Year}
\revised{Day Month Year}
\end{history}

\begin{abstract}
It was  argued in a number of papers that the gravitational
potential calculated by using the modified QFT that follows from the
Planck-length deformed uncertainty relation implies the existence of
black-hole remnants of the order of the Planck-mass. Usually this sort of QFTs 
are endowed with two specific features, the modified dispersion relation, 
which is universal, and the concept of minimum length, which, however, is not 
universal. While the emergence of the minimum-length most readily leads to the 
idea of the black hole remnants, here we examine the behaviour of the potential
that follows from the Planck-length deformed QFT in absence of the minimum 
length and show that it might also lead  to the formation of the Planck mass 
black holes in some particular cases. The calculations are made for 
higher-dimensional case as well. Such black hole remnants might be considered 
as a possible candidates for the dark-matter.              
\end{abstract}


\ccode{04.60.-m, 04.70.Dy, 04.50.Gh}

\section{Introduction}

The body of this paper is devoted to the further consideration of the black 
hole (BH) remnants in the framework of Planck-length deformed QFT put forward 
in \cite{Maziashvili:2011dx, Maziashvili:2012gm}. Before plunging into this 
analysis, though, we would like to make some clarifying comments. Within the 
framework of general relativity theory, when quantum effects are taken into 
account, black hole becomes unstable and its mass decreases as a result of 
Hawking emission \cite{Hawking:1974sw} predicting the complete vanishing of 
the black hole \cite{Hawking:1974rv}. One usually expects that the formalism 
used for describing the creation of particles by a gravitational field of the 
black hole breaks down whereas the size of the evaporating black hole becomes 
sufficiently small. The two well known arguments are as follows: 1) in the 
vicinity of the singularity at the black hole center general relativity 
becomes invalid and 2) at the Planck scale $l_P\approx 10^{-33}$cm, quantum 
fluctuations of the metric become of the order of unity and therefore the 
general relativistic description of the gravitational field no longer holds
\cite{Wheeler:1957mu, GKL}. Quantum-gravitational corrections are very 
plausibly of a size to essentially alter the Planck-length-size black hole's 
radiation. This point was recognized almost immediately on the publication of 
Hawking's work \cite{Ginzburg:1975ee}. Apart from particular theories of 
quantum gravity, there is a simple phenomenological approach to "understanding"
 of quantum-gravitational effects based on the modified uncertainty relation. 
The modification is understood to be due to quantum-gravitational fluctuations 
of the background metric. On the quite general grounds, the quantum 
fluctuations in the geometry of space can be parametrized as $\delta l \simeq 
l_P^{\alpha}l^{1-\alpha}$ on account of the mutual relation between the metric 
and the distance. This additional inaccuracy cannot be wiped out by the 
quantum mechanics; it can only increase the position uncertainty of the 
particle, not diminish it. It is straightforward to write down the modified 
uncertainty relation taking into account this additional inaccuracy due to 
quantum fluctuations of the background metric \cite{Maziashvili:2012dd} 

\begin{equation} \delta X \delta P \,\geq \, \frac{1}{2} \,+\, \beta l_P^\alpha
\delta P^\alpha ~, \label{modganuzghvrelobistanap} \end{equation} 
where $\beta$ 
stands for a numerical factor of order unity. The deformed quantum mechanics 
used throughout the discussion is motivated by this sort of uncertainty 
relation. According to this modified uncertainty relation, if $\alpha > 1$ the 
position uncertainty has its least value of the order of the Planck length. On 
the other hand, if one asks for the minimum position uncertainty in the case 
$0 < \alpha <1$, one finds no nonzero lower bound on it. Accordingly, we 
distinct those two cases by saying that in the first case there is a minimum 
length although in both cases the deformed quantum mechanics depends upon the 
Planck length. An interested reader is referred to the last section of paper 
\cite{Maziashvili:2012zr} for further details about the distinguishing between 
the minimum length and the deformation of theory. The following historical 
comment regarding the modified uncertainty relation may be helpful to orient 
the reader. The modified uncertainty relation was originally proposed purely 
in the context of high energy physics \cite{Wataghin, Saavedra:1979gc, 
SaavedraUtreras, Saavedra:1980ss, Giffon:1983bg, Talukdar:1982qc,  
Montecinos:1985zs}, however, it became popular after its "derivation" in the 
context of string theory \cite{Veneziano:1986zf, Gross:1987ar, Amati:1988tn, 
Veneziano:1989fc, Konishi:1989wk, Guida:1990st, Veneziano:1989ti, Witten:1996, 
Duff:2001ba} as a simple tool for "understanding" quantum gravitational 
effects. 

As it is stated in the abstract, the specific goal of the present paper is to 
demonstrate that the existence of black hole remnants in the framework of the 
discussion \cite{Maziashvili:2011dx, Maziashvili:2012gm} does not necessarily 
require the presence of the minimum length (see the above definition of the 
minimum length). Before proceeding to our analysis, let us briefly discuss the 
relationship between the behaviour of the gravitational potential and the 
black hole remnants. In what follows we will adopt the natural system of units 
$c=\hbar=1$. The starting point is the modified Schwarzschild (-Tangherlini) 
space-time \cite{Tangherlini:1963bw, Myers:1986un} 
 
\begin{eqnarray}\label{Schwarzschild}  
ds^2 \,=\, \left[ 1 \,-\, R_g^{n+1} V(r) 
\right]dt^2 \,-\,    \left[ 1 \,-\, R_g^{n+1}V(r) \right]^{-1}dr^2 \,-\, r^2 
d\Omega^2_{n+2} ~, \end{eqnarray}  
where $d\Omega^2_{n+2}$ is a line element of a $2+n$ dimensional unit sphere,
\begin{eqnarray}\label{standardgradius}  R_g\left(M\right) \,=\, \left(
\mathbb{G}_N M \right)^\frac{1}{n+1} \left[\frac{16 \pi }{(n+2)\text{Vol}
\left(S^{n+2}\right)} \right]^\frac{1}{n+1} ~, \end{eqnarray}

\noindent and the potential $V(r)$ is calculated by using the modified 
propagator that follows from the Planck-length deformed QFT. Essentially, the 
existence of the zero-temperature black hole remnants in the framework of this 
discussion is based on the following facts. The potential appears to be 
a monotonically decreasing function, finite at the origin with vanishing 
derivative at this point, that is: $V'(r) < 0$ for $r > 0$; $ \, V(0) < 
\infty, \, V'(0) = 0$. To see how these conditions provide the black hole 
remnant, let us notice that, in view of the Eq.\eqref{Schwarzschild}, 
the gravitational radius $r_g$ turns out to be the solution of the equation: 
\begin{equation}
V(r_g)=\frac{1}{R_g^{n+1}}=\frac{(n+2)\text{Vol}\left(S^{n+2}\right)}
{16\pi\mathbb{G}_N M}.
\label{r_g}
\end{equation}
However, if the potential is a monotonically decreasing function having its 
maximum at the origin, this equation does not have a solution for 
$$M<\frac{(n+2)\text{Vol}\left(S^{n+2}\right)}{16\pi\mathbb{G}_N V(0)}.$$
Thus, one infers that as soon as a black hole mass drops to 
$$M_{remnant}= \frac{(n+2)\text{Vol}\left(S^{n+2}\right)}
{16\pi\mathbb{G}_N V(0)},$$
the black hole horizon disappears and, on the other hand, its temperature, for 
it is proportional to the surface gravity: $V'(0)$, vanishes. It is worth 
noticing that typically, the mass of such remnants are of the order of the 
quantum gravity scale. 

Even in presence of the minimum length it is not self-evident why the 
potential estimated through the modified propagator should behave this way 
but, in this case, it completes well the intuitive picture that follows simply 
from the Poisson's equation: $\Delta \Phi = 4\pi \mathbb{G}_N\rho$. The 
presence of the non-zero minimum position uncertainty in QM engenders the 
smearing of the delta function $\rho=M\delta(\mathbf{r})$ thus replacing the 
point-like source with a regular distribution. So, in presence of the minimum 
length, the result can be recognized as an over-all picture that follows from 
the implementation of the minimum length into quantum theory 
\cite{Nicolini:2005vd, Ansoldi:2008jw, Spallucci:2011rn, Modesto:2010uh, 
Sprenger:2012uc}.

In what follows we will use the Hilbert space representation for 
a Planck-length deformed QM that explicitly depends upon the parameter 
$\alpha$ yielding the minimum length for $\alpha >1$ and leading just to the 
modified dispersion relation whereas $0 < \alpha < 1$ - as it follows from 
Eq.\eqref{modganuzghvrelobistanap}; for technicalities see 
\cite{Maziashvili:2012dd}. This sort of QM is based on the generalized 
position and momentum operators that have the following form \begin{eqnarray}
\label{pdefqmrep}  \widehat{X}_i \, = \, \widehat{x}_i ~, ~~~  \widehat{P}_j 
\,=\,  \widehat{p}_j \left(1 \,-\,  \frac{2\beta(\alpha-1)}{\alpha} \, 
l_P^\alpha \widehat{p}^\alpha  \right)^{\frac{1}{1-\alpha}} ~, \end{eqnarray} 
where 
$\widehat{\mathbf{x}}, \, \widehat{\mathbf{p}}$ stand for the standard position
and momentum operators in the $p$-representation, $\beta$ is 
a numerical factor of order unity and $l_P$ denotes the Planck length.
In the case $\alpha=2$, the Eq.\eqref{pdefqmrep} reduces to the well-known 
result found in \cite{Kempf:1996nk}. Let us notice that in view of 
Eq.\eqref{pdefqmrep}, there is a cutoff $p^{\alpha} \,<\, \alpha/2\beta(\alpha
-1)l_P^\alpha$ when $\alpha > 1$. This cutoff arises merely from the fact that 
when small $p$ runs over this region, large $P$ covers the whole region from 
$0$ to $\infty$ (for more details see \cite{Maziashvili:2012dd}). Indeed this 
cutoff is responsible for the existence of the non-zero  minimum position 
uncertainty. In the case $0< \alpha < 1$ there is no lower bound on position 
uncertainty and no cutoff on $p$, respectively.

In light of the Eq.\eqref{pdefqmrep}, the dispersion relation for a free 
massless particle and the correspondingly modified field theory action read
\begin{eqnarray}\label{PlmodQFT} \varepsilon = P^2~,~~~~ \mathcal{A}[\varPhi] =
 - \int d^4x \, \frac{1}{2} \left[\varPhi\partial_t^2\varPhi  + \varPhi{\mathbf
 P}^2\varPhi \right]  ~.\end{eqnarray} In what follows, we will use the 
propagator following from Planck-length deformed field theory \eqref{PlmodQFT} 
to estimate the modified potential, which then will be used in 
Eq.\eqref{Schwarzschild}. To be more precise, the field theory model for 
gravity we are dealing with is obtained by expanding the GR around the 
Minkowskian metric and then substituting $\partial_i=i\widehat{p}_i$ by 
the non-local operator $i\widehat{P}_i$, as it is done in Eq\eqref{PlmodQFT}. 
So, this theory contains an infinite number of higher derivative terms. 
What we know for sure about this theory is that one can use the weak field 
approximation in the large distance limit and in this limit the theory admits 
a BH-type soliton solution, so to say \cite{Salam:1976mm}. The new feature 
that follows from the calculations made here is that for $\alpha >(2+n)/(4+n)$ 
this theory admits the weak field approximation in the short distance limit 
as well. So, strictly speaking the modified BH solution we are discussing is 
valid just in these asymptotic regions, however, we hope, the conclusion about 
the BH remnants is not affected by this fact. Our purpose, as stated in the 
manuscript, was to complete the discussion of \cite{Nicolini:2005vd, 
Ansoldi:2008jw, Spallucci:2011rn}, where the effect of minimum length on the 
matter fields is discussed in the context of BH physics. But we are discussing 
somewhat less intuitive case in which there is no minimum length as such, that 
is, position uncertainty has no non-zero bound from below, or otherwise 
speaking, there is no cut-off on p, and therefore the arguments of 
\cite{Nicolini:2005vd, Ansoldi:2008jw, Spallucci:2011rn} simply do not work.

\section{The behaviour of the potential}

\subsection{No minimum length: $0< \alpha < 1$}

From the above discussion it follows immediately that the modified Poisson 
equation for the point-like source takes the form 

\begin{equation}\label{modPoissongantoleba}
\hat{P}^2  V(\mathbf{r}) \, =\, 4 \pi  \delta(\mathbf{r}) ~, 
\end{equation} so that the potential calculated by using the modified
propagator that follows from Eqs.(\ref{pdefqmrep}, \ref{PlmodQFT}), which is 
nothing else but the solution of Eq.\eqref{modPoissongantoleba}, takes the 
form 

\begin{eqnarray}\label{potrep}
V(r) \,=\, \int \frac{d^3k}{(2\pi)^3} \, \frac{4\pi e^{i\mathbf{k}\mathbf{r}}}
{k^2 \left(1 + \mbox{\ss} k^\alpha \right)^{2/(1-\alpha)}} ~,
\end{eqnarray} where 
\[ \mbox{\ss} \,\equiv\, \frac{2\beta(1 \,-\, \alpha)l_P^\alpha}{\alpha} ~. \]

The Fourier integrals in higher-dimensional case are more divergent but the 
basic structure is essentially the same: 
\begin{equation}\label{maghalganzpot}
V(r) = \frac{(1+n)\mathrm{Vol}\left(S^{n+2}\right)}{(2\pi)^{3+n}}\int  
\frac{d^{3+n}k \, \, e^{i\mathbf{k}\cdot\mathbf{r}}}{k^2 \left(1 + \mbox{\ss} 
k^\alpha \right)^{2/(1-\alpha)}},
\end{equation} 
where $n$ denotes the number of extra dimensions. From this expression one 
finds that $V(0)$ is finite as long as $\alpha > (1+n)/(3+n)$. To work out 
the behaviour of the potential for $r\rightarrow 0$, let us write the 
Eq.\ref{maghalganzpot} in the form 
\begin{equation}\label{maghalganzpotyofnul}
V(r)= 
\frac{K}{r^{n+1}} \int\limits_{-1}^1 d\tau \, 
\left(1-\tau^2\right)^\frac{n}{2}\int\limits_0^\infty  \frac{dq \, q^{n}
\cos(q\tau)}{ \left[1 + \mbox{\ss} (q/r)^\alpha \right]^{2/(1-\alpha)}},
\end{equation}
where
\[ K=\frac{(1+n) \,\mathrm{Vol}\left(S^{n+2}\right)\mathrm{Vol}\left(S^{n+1}
\right)}{(2\pi)^{3+n}} ~, \]
and $q$ and $\tau$ are dimensionless quantities defined through 
$q=|\mathbf{k}|r$ and $\mathbf{k}\cdot\mathbf{r}=q\tau$.

First of all, let us notice that the integral with respect to $q$ in 
Eq.\ref{maghalganzpotyofnul} is understood in the sense of the generalized 
functions (for a detailed account of generalized functions/distributions we 
refer the reader to the review article \cite{Guttinger}). It may be divergent 
for some values of $\alpha$ (namely, if $2/(1-\alpha)<n+1$) but nevertheless 
the second integral with respect to $\tau$ gives the finite result (see the 
discussion after Eq.\ref{potentstsarmoebulimganshem}). Keeping this in mind, 
one can split the integral       
\begin{eqnarray} &
\int\limits_{-1}^1 d\tau \, \left(1-\tau^2\right)^\frac{n}{2} 
\int\limits_0^\infty  \frac{dq \, q^{n}\cos(q\tau)}{ \left[1 + \mbox{\ss} 
(q/r)^\alpha \right]^{2/(1-\alpha)}} \,=  \int\limits_{-1}^1 d\tau \, 
\left(1-\tau^2\right)^\frac{n}{2}& \nonumber \\ & \times \,\left\{  \int
\limits_0^\epsilon  \frac{dq \, q^{n}\cos(q\tau)}{ \left[1 + \mbox{\ss} 
(q/r)^\alpha \right]^{2/(1-\alpha)}}  \,+     \int\limits_\epsilon^\infty  
\frac{dq \, q^{n}\cos(q\tau)}{ \left[1 + \mbox{\ss} (q/r)^\alpha 
\right]^{2/(1-\alpha)}} \right\},&\nonumber
\end{eqnarray} 
($\epsilon\ll 1$) and omit the second term as it goes to zero when 
$r \rightarrow 0$. Thus, one infers that the short distance behaviour 
$\left(r \ll \epsilon\mbox{\ss}^{1/\alpha}\right)$ of the potential 
is essentially determined by the integral 
\begin{eqnarray}\label{esetsdaskvnashi}
V\left(r\ll \epsilon\mbox{\ss}^{1/\alpha}\right) \,\simeq\, 
~~~~~~~~~~~~~~~~~~~~~~~~~~~~~~~~~~\nonumber\\ \frac{(1+n)\mathrm{Vol}
\left(S^{n+2}\right)}{(2\pi)^\frac{3+n}{2}\, 2^\frac{n+1}{2}\Gamma\left(
\frac{n+3}{2}\right)}\int\limits_0^{\epsilon/r}  \frac{dk \, \, k^n}{ 
\left[1 + \mbox{\ss} k^\alpha \right]^{2/(1-\alpha)}}   ~,
\end{eqnarray} 
where we have used 
\[
\int\limits_{-1}^1 d\tau \, \left(1-\tau^2\right)^\frac{n}{2} \,=\, 
\frac{\Gamma\left(\frac{1}{2}\right)\Gamma\left(1+\frac{n}{2}\right)}
{\Gamma\left(\frac{3+n}{2}\right)} ~. \]
Correspondingly, one finds 
\begin{eqnarray}\label{daesets}
V'\left(r\ll \epsilon\mbox{\ss}^{1/\alpha}\right) \,\propto\, -\, 
r^\frac{(4+n)
\alpha-n-2}{1-\alpha}  ~,
\end{eqnarray} 
that is, $V'(0)=0$ when $\alpha > (2+n)/(4+n)$. This result can be seen 
immediately from the Eq.\ref{maghalganzpot}. 

One can show that in general $V'(r)< 0$ for $r>0$, that is, $V(r)$ is 
a monotonically decreasing function for $r>0$. Namely, the derivative of 
the potential reads  
\begin{eqnarray}\label{potentstsarmoebulimganshem}
V'(r) \,=\, \nabla V(r) \cdot \frac{\mathbf{r}}{r} \,=\,
-\, \frac{K}{r^{n+2}} \int\limits_{-1}^1 d\tau \, \left(1-\tau^2\right)
^\frac{n}{2}\tau \nonumber \\
\times\int\limits_0^\infty  \frac{dq \, q^{n+1}\sin(q\tau)}
{ \left[1 + \mbox{\ss} (q/r)^\alpha \right]^{2/(1-\alpha)}}. 
\end{eqnarray}
As in the previous case, it can be seen that the integral 
\[
\int\limits_0^\infty  \frac{dq \, q^{n+1}\sin(q\tau)}{ \left[1 + \mbox{\ss} 
(q/r)^\alpha \right]^{2/(1-\alpha)}} \,\equiv\,\Im  \int\limits_0^\infty  
\frac{dq \, q^{n+1}e^{iq\tau}}{ \left[1 + \mbox{\ss} (q/r)^\alpha 
\right]^{2/(1-\alpha)}} ~, \]
that enters the Eq.\ref{potentstsarmoebulimganshem} may be divergent for 
certain values of $\alpha$ (for a given $n$) but still integrable with 
respect to $\tau$. Again, this integral should be understood by introducing 
the factor $e^{-\varepsilon q }$ in the integral and taking the limit 
$\varepsilon\to 0$ afterwords. So, one can use the following relation 
\cite{BogolyubovShirkov} ($\mathcal{P}$ stands for the principal value) 
\begin{eqnarray}
\int\limits_0^\infty dq \, e^{iq(\tau + i\varepsilon)} 
\to \pi\delta(\tau) \,+\, \mathcal{P}\frac{i}{\tau} ~.
\end{eqnarray} 
Since in the standard case, $\mbox{\ss} =0 $, the double integral entering 
the Eq.\ref{potentstsarmoebulimganshem} is positive, then by taking into 
account that the integrand is now divided by the monotonically increasing 
function, one concludes that its positiveness is guaranteed. The argument 
is that one can write the integral with respect to $\tau$ over the region 
$(0, 1)$, that is, with the integration starting from zero. The extra factor 
on which the sine-function is multiplied is monotonically decreasing in this 
case, so the negative contribution coming from this function in the integral 
is now more suppressed as compared to its positive contribution. Considering 
the worse case, $\mbox{\ss}=0$, one can calculate the 
(\ref{potentstsarmoebulimganshem}) straightforwardly with the use of equation
\begin{eqnarray}\label{ganzogadtsarmoebuli}
\int\limits_0^\infty dq \, q^{n+1} e^{iq\tau} \,=\, (-i)^{n+1} \, 
\frac{\partial^{n+1}}{\partial\tau^{n+1}}\int\limits_0^\infty e^{iq\tau} 
\,= \,\nonumber \\
(-i)^{n+1} \, \pi \, \delta^{(n+1)}(\tau) \,-\, (-1)^n(-i)^{n}\, (n+1)!\, 
\mathcal{P} \, \frac{1}{\tau^{n+2}} ~, ~~
\end{eqnarray} 
where the principal value $\mathcal{P}$ is understood in the sense of 
Hadamard \cite{Fox} 
\begin{eqnarray}\label{HadamardFox}
&&\mathcal{P}\int\limits_a^b dx\, \frac{f(x)}{(x-u)^{n+1}} \,=  \nonumber \\ 
&&\frac{f(a)}{n(a-u)^n} \,-\, \frac{f(b)}{n(b-u)^n} \,+\, \frac{1}{n}\, 
\mathcal{P} \int\limits_a^b dx \, \frac{f^{(1)}(x)}{(x-u)^n} ~. ~~~~
\end{eqnarray} 
Instead of using Eqs.(\ref{ganzogadtsarmoebuli}, \ref{HadamardFox}) 
straightforwardly, one could calculate this type of integrals by using the 
$\epsilon$ prescription but this approach is somewhat more cumbersome ---  
see the Appendix. Using the Eqs.(\ref{ganzogadtsarmoebuli}, 
\ref{HadamardFox}), one finds that in the case $\mbox{\ss}=0$ the derivative 
of the potential (\ref{potentstsarmoebulimganshem}) takes the form 
\begin{eqnarray}
V'(r)\,=\,- \frac{K\pi\,(n+1)!}{r^{n+2}}.
\end{eqnarray}

\begin{figure}[ht]
\includegraphics[width=0.8\textwidth]{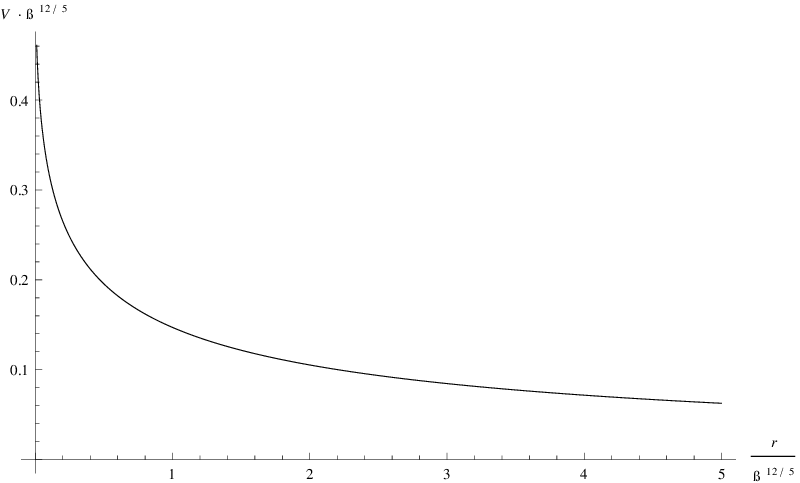}\\
\caption{The typical behaviour of the potential for $1/3 < \alpha < 1/2$. The 
essential properties are: $V(0) <\infty, V'(r) <0$ for $r>0$ and $V'(r) 
\rightarrow -\infty$ as $r\rightarrow 0$. (The plot is made in particular for 
the case $\alpha=5/12$).}
\label{Abbildung1}
\end{figure}

\begin{figure}[ht]
\includegraphics[width=0.8\textwidth]{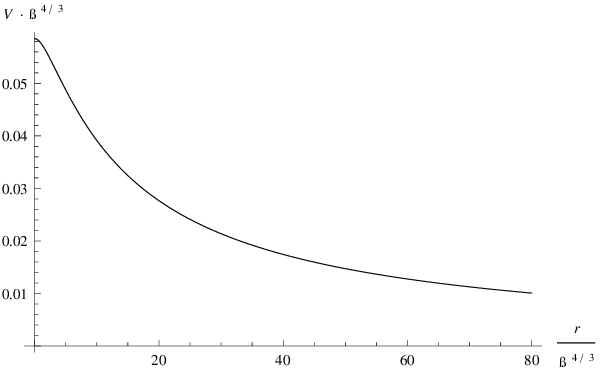}\\
\caption{The typical behaviour of the potential for $1/2 < \alpha < 1$. The 
essential characteristics are: $V(0) <\infty, V'(r) <0$ for $r>0$ and $V'(r) 
\rightarrow 0$ as $r\rightarrow 0$. (The plot is made in particular for the 
case $\alpha=3/4$).}
\label{Abbildung2}
\end{figure}

The typical behaviours of the potential in three-dimensional case are  shown
in Fig.\ref{Abbildung1} for $1/3 < \alpha < 1/2$ and in Fig.\ref{Abbildung2}
for  $1/3 < \alpha < 1/2$.

\subsection{Minimum length: $\alpha > 1$}
Because of the presence of cut-off $k < \mbox{\ss}^{-1/\alpha}$ in the case 
of minimum length, one can readily show all of the features required for the 
potential for the existence of the zero-temperature BH remnants. The potential 
looks like   
\begin{eqnarray}\label{maghalganzpotyofnul1}
V(r) \,=\, K  
\int\limits_0^{\mbox{\ss}^{-1/\alpha}} dk \, k^{n} \left(1 - \mbox{\ss} 
k^\alpha \right)^{2/(\alpha-1)} \nonumber \\ \times 
\int\limits_{-1}^1 d\tau \, \left(1-\tau^2\right)^\frac{n}{2} \cos(kr\tau).
\end{eqnarray}
It is evident from this expression that $V(0)$ is finite, $V'(0)=0$ and 
$V'(r) < 0$. The last statement can be proved much in the same way as it was 
done in the case $0< \alpha < 1$. As we see, the potential has the same 
typical behaviour as  shown in Fig.\ref{Abbildung2}. The asymptotic behaviour 
of $V(r)$ for $r \rightarrow 0$ can easily be found as well by expanding the 
$\cos\left(krt\right)$  into the Taylor series. One finds 
\begin{equation}\label{potmravalchamyofsatav}
V(r) \,=\, \mathcal{A} \,-\, \mathcal{B}r^2 \,+\, O(r^4)~,
\end{equation} where $\mathcal{A}$ and $\mathcal{B}$ are positive quantities.

\section{Summary and discussion}

Main goal of this paper was to present more or less systematic account of how 
modified uncertainty relations of the sort given by 
Eq.\eqref{modganuzghvrelobistanap} might affect the black hole physics. Two 
kind of modifications suggested by the deformed uncertainty relations that 
should be distinguished are as follows. The primary and unique feature of it 
is that it implies modified dispersion relation and the second, non-universal 
feature, is the appearance of the lower non-zero bound on the position 
uncertainty (for which we saved the term - minimum length). The latter feature 
most readily leads to the idea of the BH remnants as it implies the smearing 
of the matter fields over the region of size $l_P$ - ruling out the point-like 
sources. To quantify, this feature is mathematically expressed in emergence of 
the cutoff on the standard momentum variable (which is canonically conjugate 
to the coordinate). But one should take into account that this cutoff affects 
not only matter fields but the gravitational field as well. Interestingly 
enough, by taking account this cutoff just for the gravitational field, one 
arrives at the same qualitative results that follow from the theory augmented 
by the smearing out of the matter fields \cite{Maziashvili:2011dx}. However, 
putting aside the latter feature of modified uncertainty relation, merely the 
deformation of the dispersion relation can also lead to the halt of the BH 
emission much in the same way as it is achieved by exploiting the concept of 
the minimum length. That is what we have shown throughout this paper in the 
framework of a somewhat generic setup. An appropriate question that naturally 
arises in absence of the minimum length, is the remnant's size. We have seen 
that the horizon goes to zero when mass approaches the Planck scale but BH 
implies the size of the object should be smaller than its gravitational radius.
Does it mean that it's a point-like object? The answer is yes, because the 
very approach we are pursuing starts merely from the Poisson equation with the 
point-like source.     

Now let us summarize the results in some detail (we will restrict ourselves 
just to the 3D case). The parameter range $0 < \alpha < 1$ implies the absence 
of the minimum length. The Planck-length deformed propagator for $\alpha > 1/3$
results in the potential, which after being used in the Schwarzschild metric 
shows up the existence of the black hole remnants (disappearance of horizon), 
however, for values $1/3 < \alpha <1/2$ the temperature of the black hole 
remnant goes to infinity. In contrast, for $\alpha > 1/2$ the black hole 
remnants are characterized with the zero temperature (The typical behaviour 
of the radiation temperature as a function of the BH mass is shown in Fig.1). 
Moreover, that modified Schwarzschild space-time is free of the curvature 
singularity at the origin when $\alpha > 3/5$. That is easy to see for in this 
case the metric as well as its first and second derivatives do not diverge 
when $r\rightarrow 0$ (see Eqs.(\ref{esetsdaskvnashi}, \ref{daesets}, 
\ref{potmravalchamyofsatav})). Let us emphasize once again that the key 
observation made throughout this paper is the existence of the black hole 
remnants when there is no minimum length: $\alpha < 1$. 

\begin{figure}[ht]

\includegraphics[width=0.8\textwidth]{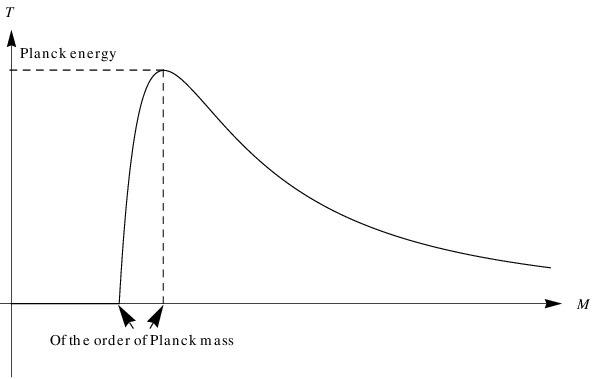}\\
\caption{ Typical behaviour of the emission temperature as a function of the 
BH mass for $\alpha > (2+n)/(4+n)$. The emission temperature reaches its 
maximum - of the order of Planck energy, when BH evaporates down to the Planck 
mass, then it swiftly drops to zero at $M_{remnant}$, which also is of the 
order of Planck mass.  }
\end{figure} 

Special attention has to be paid on the validity conditions of approximation 
assumed tacitly throughout the above discussion. We have taken gravitational 
field on the equal footing with the matter fields, that is, QFT picture for 
gravity is taken as a starting point. This means that the graviton field is 
defined as the difference between the full metric and its Minkowski 
background value and a field theory on flat Minkowski spacetime is assumed 
to hold for this graviton field. Such QFT approach to gravity, pioneered by 
Kraichnan (the only post-doctoral student that Einstein ever had) 
\cite{Kraichnan:1955} and Gupta \cite{Gupta:1954}, is reviewed in 
\cite{Feynman:1995,Thirring:1961,Straumann:2000,Padmanabhan:2004}.
However, we have used this QFT approach to gravity, suitably modified, only 
to get the generalization of the Newton potential and then embarked on 
more traditional geometric approach by substituting this generalized
potential into the Schwarzschild-Tangherlini metric. Although quite 
reasonable in a weak-field limit, it seems completely impossible to justify 
the use of this substitution up to the Planck scale. Nevertheless, the
following rather ingenious, though somewhat heuristic, argument can be 
envisaged to justify such kind of business. We will assume $n=0$ (that is 
3D case) in the following.

It is relatively little known that the complete content of Einstein's 
general relativity is encoded in the following single equation 
\cite{Feynman:1995,Santander:1997} ($c=1$ is assumed as earlier):
\begin{equation}
\mathcal{K}(12)+\mathcal{K}(23)+\mathcal{K}(31)= 8\pi\mathbb{G}_N W^0,
\label{Kequation}
\end{equation}
valid irrespective of the state of motion of the observer. Here $W^0$
is the proper energy density at the considered space-time point measured
in the proper comoving frame and $\mathcal{K}(ij)$ are sectional curvatures 
in the proper three space. Of course, this single equation valid for any 
observer implies a set of equations which should hold true for each observer
and this set of equations are equivalent to usual tensorial form of 
Einstein's equations \cite{Santander:1997}. In the Newtonian limit 
$c\to\infty$, one of these equations reduces to \cite{Santander:1997}
$$\mathcal{K}(01)+\mathcal{K}(02)+\mathcal{K}(03)= 4\pi\mathbb{G}_N \rho,$$
which is exactly the Poisson equation for the Newtonian gravitational 
potential $V$ because in this limit 
$$\mathcal{K}(01)=\frac{\partial^2 V}{\partial x^2},\;\;
\mathcal{K}(02)=\frac{\partial^2 V}{\partial y^2},\;\;
\mathcal{K}(03)=\frac{\partial^2 V}{\partial z^2}.$$
This fact explains why we have chosen a modification of the Poisson equation,
inspired by Planck-length deformed QFT, as our starting point. Of course 
it is assumed that this modification of the Poisson equation is just 
a limiting case of suitably modified gravitational field equations. In the 
following we conjecture one such modification.

Namely, let us consider the following non-local modification of the equation 
(\ref{Kequation}) (and similar modifications of its accompanying equations),
inspired by the Planck-length deformed QM as given by relations 
(\ref{pdefqmrep}),
\begin{equation}
\widehat{L}\,(\,\mathcal{K}(12)+\mathcal{K}(23)+\mathcal{K}(31)\,)= 
8\pi\mathbb{G}_N W^0,
\label{MKequation}
\end{equation} 
where
\begin{equation}
\widehat{L}=\left(1 \,-\,\mbox{\ss}\widehat{p}^{\,\alpha}  
\right)^{\frac{2}{1-\alpha}},
\label{Loperator}
\end{equation}
and $\widehat{p}=\sqrt{-\Delta}$.

In the spherically symmetric three space around a point-like mass distribution
$W^0=M\delta(\mathbf{r})$ we have $\mathcal{K}(\theta r)=\mathcal{K}
(\phi r)=D(r)$ and $\mathcal{K}(\theta\phi)=T(r)$ with two unknown functions 
$D$ and $T$ \cite{Santander:1997}. Therefore, according to
(\ref{MKequation}), in the empty space, beyond the origin, we should have
\begin{equation}
\widehat{L}(2D+T)=0.
\label{L2DT}
\end{equation}
On the other hand, in general relativity the functions $D$ and $T$ are related 
due to Bianci identities in the following way \cite{Santander:1997}:
\begin{equation}
r\frac{dT}{dr}=2(D-T).
\label{dTdr}
\end{equation}
It turns out that self-consistency of our deformation of Schwarz\-schild 
solution requires the following modification of (\ref{dTdr}):
\begin{equation}
\widehat{L_1}\left (r\frac{dT}{dr}+3T\right )=\widehat{L}(2D+T),
\label{dTdrM}
\end{equation}
where $L_1=(1+\mathbf{r}\cdot\mathbf{\nabla})\widehat{L}(1+\mathbf{r}\cdot
\mathbf{\nabla})^{-1}$.

Combining equations (\ref{L2DT}) and (\ref{dTdrM}), we get
\begin{equation}
\widehat{L_1}\left (r\frac{dT}{dr}+3T\right)=\widehat{L_1}\left (\frac{1}{r^2}
\frac{d}{dr}(r^3T)\right)=0,
\label{Tmain1}
\end{equation}
Noticing that $1+\mathbf{r}\cdot\mathbf{\nabla}$, and hence $(1+\mathbf{r}
\cdot\mathbf{\nabla})^{-1}$, commutes with $3+\mathbf{r}\cdot\mathbf{\nabla}$,
we can rewrite (\ref{Tmain1}) in the form
\begin{equation}
(1+\mathbf{r}\cdot\mathbf{\nabla})\left(1 \,-\,  
\mbox{\ss} \widehat{p}^{\,\alpha}  
\right)^{\frac{2}{1-\alpha}}\left[\frac{1}{r^2}\frac{d}{dr}\left(r^3
F\right)\right]=0,
\label{Tmain2}
\end{equation}
where $F=(1+\mathbf{r}\cdot\mathbf{\nabla})^{-1}T$.

On the other hand, in the same empty space-region our potential $V(r)$ 
satisfies the modified Poisson equation
$$\hat{P}^2  V(r)=-\left(1 \,-\,  
\mbox{\ss} \widehat{p}^{\,\alpha}  
\right)^{\frac{2}{1-\alpha}}\Delta V=0.$$
Therefore, we get a solution of (\ref{Tmain2}) if we take
$$\frac{1}{r^2}\frac{d}{dr}\left(r^3 F\right)=k\Delta V=\frac{k}{r^2}
\frac{d}{dr}\left(r^2\frac{dV}{dr}\right ),$$
with some constant $k$, which implies
$$(1+\mathbf{r}\cdot\mathbf{\nabla})^{-1}T=\frac{k}{r}\frac{dV}{dr},$$
and
\begin{equation}
T(r)=(1+\mathbf{r}\cdot\mathbf{\nabla})\left(\frac{k}{r}\frac{dV}{dr}\right)
=k\frac{d^2 V}{dr^2}.
\label{Tfinal}
\end{equation}
It is well known that in the case of static, spherically symmetric 
distribution of matter, without loss of generality, we can assume the 
Schwarzschild-like form for metric in the outside region with two unknown 
functions \cite{Tangherlini:1962}
\begin{equation}
ds^2=f(r)dt^2-g(r)dr^2-r^2 d\Omega^2_2.
\label{ds2general}
\end{equation}
The radial geodesic equation which follows from this metric has the form
\cite{Tangherlini:1962,Rindler:1969}
\begin{equation}
\frac{d^2r}{ds^2}=\frac{1}{2}\frac{d}{dr}\left(\frac{k_0^2}{fg}-\frac{1}{g}
\right),
\label{Rgeodesic}
\end{equation}
where $k_0$ is the energy (per unit mass) of the test particle. In Newtonian
gravity the radial acceleration depends merely on the gradient  of the the 
gravitational potential at the location of the particle. The generalization
of this characteristic property of the Newtonian gravity, which Tangherlini
calls strong principle of equivalence \cite{Tangherlini:1962},
implies that $f(r)g(r)$ is constant to eliminate the $k_0$-dependent term in
(\ref{Rgeodesic}). When, as the space-time is assumed to be asymptotically
Minkowskian, we get $g(r)=1/f(r)$. In this case the metric (\ref{ds2general})
implies the following sectional curvature \cite{Santander:1997}
\begin{equation}
\mathcal{K}(tr)=\frac{1}{4fg}\left[2\frac{d^2f}{dr^2}-\frac{1}{f}\left(
\frac{df}{dr}\right)-\frac{1}{g}\left(\frac{df}{dr}\right)\left(\frac{dg}{dr}
\right)\right]=\frac{1}{2}\frac{d^2f}{dr^2}.
\label{Ktr}
\end{equation}
On the other hand one of accompanying equations of (\ref{Kequation}) implies
that in the standard case $\mathcal{K}(tr)=-\mathcal{K}(\theta\phi)=-T(r)$ 
\cite{Santander:1997}. 
This remains a possible solution in the modified case also when the operator
$\widehat{L}$ is assumed to act only on sectional curvatures and not on matter 
source terms. Therefore,
$$\frac{1}{2}\frac{d^2f}{dr^2}=-k\frac{d^2 V}{dr^2},$$
whose solution, with the proper asymptotic, is
\begin{equation}
f(r)=1-2kV(r)=1-R_gV(r),
\label{frfinal}
\end{equation}
where the constant $k$ was fixed by requiring the proper (that is 
Schwarzschild) limit in the case $\beta\to 0$.

The aim of this long argument was to demonstrate that we can anticipate such 
Planck-length deformation of Einstein's equations for which 
(\ref{Schwarzschild}), with $V(r)$ given by  (\ref{maghalganzpot}), is an 
exact and not merely an approximate solution, much like the standard 
Schwarzschild case.

We do not know a geometric meaning of the modification of Bianci identities
(\ref{dTdrM}). The necessity of this modification is caused by the fact 
that without it we get a contradiction if in the sectional curvature 
\cite{Santander:1997}
\begin{equation}
\mathcal{K}(\phi r)=-\frac{1}{2r}\frac{d}{dr}\left(\frac{1}{g}\right)=D(r)
\label{KphirD}
\end{equation}
we substitute 
$$\frac{1}{g}=1-2kV(r),\;\;\mathrm{and}\;\;D=T+\frac{1}{2}r\frac{dT}{dr},$$
with $T$ from (\ref{Tfinal}). If instead we take the modified case of 
equations (\ref{L2DT}) and (\ref{dTdrM}), the contradiction disappears.
Indeed, (\ref{KphirD}) and (\ref{L2DT}) imply
$$\widehat{L}\left(D+\frac{1}{2r}\frac{d}{dr}\left(\frac{1}{g}\right)\right)=
-\frac{1}{2}\widehat{L}\left(T-\frac{1}{r}\frac{d}{dr}\left(\frac{1}{g}
\right)\right)=0,$$
which is valid for $1/g=1-2kV$ and $T=k\frac{d^2V}{dr^2}$ because
$$-\widehat{L}\left(\frac{d^2V}{dr^2}+\frac{2}{r}\,\frac{dV}{dr}\right)=
-\widehat{L}\Delta V=\hat{P}^2  V=0.$$

The above discussed modification of the Schwarzschild metric was inspired
by papers \cite{Santander:1997} and \cite{Hokkyo:1971}. However our approach
is quite different from the one advocated in \cite{Hokkyo:1971}.

We have assumed, and our calculations of the potential are consistent with 
this assumption, that as long as $\alpha >(2+n)/(4+n)$ the 
gravity behaves as an asymptotically free interaction and, correspondingly, 
the radiative corrections close to the Planck scale can be safely ignored in 
this case. In contrast, when $\alpha < (2+n)/(4+n)$ the gravitational force 
does not go to zero when $r\rightarrow 0$ and one can not justify the 
ignorance of the radiative corrections near the Planck scale.

As a further remark, let us notice that we do not know what the full 
implementation of minimum-length deformed quantum mechanics in GR might look 
like. If one truncates the modified field theory given by Eqs.(\ref{pdefqmrep},
\ref{PlmodQFT}) to some power of $\mbox{\ss}$, then it will result in a so 
called Lifshitz like theory and therefore one might expect the corresponding 
gravity theory to look something like the Horava-Lifshitz gravity 
\cite{Horava:2009uw}. Another way of modifying the GR with respect to the 
minimum-length concept might be a non-local theory of GR 
\cite{Krasnikov:1987yj, Biswas:2011ar, 
Modesto:2011kw, Modesto:2012ys}; the question of black hole remnants in this 
sort of theory was addressed in \cite{Nicolini:2012eu}. Let us also mention 
some papers known to us addressing the question of modified potential due to 
deformed propagator \cite{Tkachuk:2007zz, Helling:2007zv, 
AmelinoCamelia:2010rm, Moayedi:2013nxa, Moayedi:2013nba} and some of the 
papers devoted to the black hole remnants due to Planck-length deformed field 
theory \cite{Adler:2001vs, Koch:2005ks, Nozari:2005ah, Nicolini:2008aj, 
Banerjee:2010sd, Koch:2008qq, Bleicher:2010qr, Nicolini:2011nz, 
Bleicher:2011uj, Nicolini:2012eu, Nicolini:2013ega}.    

It was shown that for black holes with inner (Cauchy) horizon no remnant 
formation is expected due to the so called mass inflation instability 
\cite{Poisson:1990eh, Brown:2011tv}. In our case, however, thanks to the 
properties of the potential described in the text, one can have at most one
horizon, and if the horizon is present it is much like the Schwarzschild 
case and, therefore, this type of instability does not occur. 

As a final comment, let us notice that a sufficient amount of small black
 holes can be produced in the early universe in order to consider Planck mass 
BH remnants as a viable candidates for the dark-matter. \cite{Chen:2002tu, 
Chen:2004ft, Ivanov:1994pa, GarciaBellido:1996qt, Yokoyama:1995ex, 
Kawasaki:1998vx, Kawasaki:2006zv}.

\subsection*{Appendix}

Using the $\epsilon$ prescription, in the case $\mbox{\ss}=0$ the double 
integral entering the Eq.\eqref{potentstsarmoebulimganshem}: 
\begin{eqnarray}I_n=\int\limits_{-1}^1 d\tau \, (1-\tau^2)^{n/2}\tau\int
\limits_0^\infty dq\, q^{n+1}e^{iq(\tau+i\epsilon)} \,=\, \nonumber \\ 
(-1)^{n+1}\frac{d^{n+1}}
{d\epsilon^{n+1}}\int\limits_{-1}^1d\tau \, (1-\tau^2)^{n/2}\tau\int\limits_0
^\infty dq\, e^{iq(\tau+i\epsilon)} \nonumber \end{eqnarray}
can be written in the form
\begin{eqnarray}I_n=  i(-1)^{n+1}\frac{d^{n+1}}
{d\epsilon^{n+1}}\int\limits_{-1}^1d\tau \,
(1-\tau^2)^{n/2}\frac{\tau^2+\epsilon^2-\epsilon^2}{\tau^2+\epsilon^2} \,=\,
\nonumber \\  
i(-1)^n\frac{d^{n+1}}{d\epsilon^{n+1}}\left[\epsilon^2\int\limits_{-1}^1 
d\tau \, 
\frac{(1-\tau^2)^{n/2}}{\tau^2+\epsilon^2} \right] ~. \nonumber \end{eqnarray} 
Making the substitution $\tau=\sin{\theta}$, one finds
\begin{eqnarray}
I_n=i(-1)^n\frac{d^{n+1}}{d\epsilon^{n+1}}\left[
\epsilon^2\int\limits_{-\pi/2}^{\pi/2} d\theta \, \frac{\cos^{n+1}{\theta}}
{\sin^2{\theta}+\epsilon^2}\right] \,\equiv \,
i(-1)^n\frac{d^{n+1}}{d\epsilon^{n+1}}\left[\epsilon^2 \mathcal{I}_{n+1}
(\epsilon)\right] ~. ~~~~~~
\label{eq1}
\end{eqnarray}
So, we have
\begin{eqnarray}\mathcal{I}_n(\epsilon)=\int\limits_{-\pi/2}^{\pi/2}
\frac{\cos^n{\theta}}
{\sin^2{\theta}+\epsilon^2}\;d\theta \,=\,  \int\limits_{-\pi/2}^{\pi/2}
\frac{\cos^{n-2}{\theta}(1-\sin^2{\theta}-\epsilon^2+\epsilon^2)}
{\sin^2{\theta}+\epsilon^2}\;d\theta ~,\end{eqnarray}
and hence
\begin{equation}
\mathcal{I}_n(\epsilon)=(1+\epsilon^2)\mathcal{I}_{n-2}(\epsilon)-K_{n-2},
\label{eq2}
\end{equation}
where
$$K_n=\int\limits_{-\pi/2}^{\pi/2}\cos^n{\theta}\;d\theta.$$
Using (\ref{eq2}), we can prove by induction that
\begin{equation}
\mathcal{I}_{2m}=(1+\epsilon^2)^m\mathcal{I}_0-\sum\limits_{j=0}^{m-1}(1+
\epsilon^2)^j
K_{2(m-1)-2j},
\label{eq3}
\end{equation}
and
\begin{equation}
\mathcal{I}_{2m+1}=(1+\epsilon^2)^m\mathcal{I}_1-\sum\limits_{j=0}^{m-1}(1+
\epsilon^2)^j
K_{2m-1-2j}.
\label{eq4}
\end{equation}
Hence, when $n=2m-1$ is odd, we get from Eqs.(\ref{eq1}, \ref{eq3})
\begin{equation}
I_{2m-1}=-i\frac{d^{2m}}{d\epsilon^{2m}}\left
[\epsilon^2(1+\epsilon^2)^m\mathcal{I}_0(\epsilon)-
\epsilon^2(1+\epsilon^2)^{m-1}K_0\right ],
\label{eq5}
\end{equation}
all other terms in Eq.\eqref{eq3} are giving zero contribution. Going further, 
one easily finds the values of $\mathcal{I}_0$ and $K_0$ 
$$\mathcal{I}_0(\epsilon)=\int\limits_{-\pi/2}^{\pi/2}\frac{d\theta}
{\sin^2{\theta}+\epsilon^2}=\frac{\pi}{\epsilon\sqrt{1+\epsilon^2}} 
~,~~~~ K_0=\pi ~,$$
and, therefore, the Eq.\eqref{eq5} reduces to
$$I_{2m-1}=(2m)!i\pi-i\pi\frac{d^{2m}}{d\epsilon^{2m}}\left[\epsilon
(1+\epsilon^2)^{m-1/2}\right ],$$
which is the same as
\begin{equation}
I_{2m-1}=(2m)!i\pi-\frac{i\pi}{2m+1}\frac{d^{2m+1}}{d\epsilon^{2m+1}}
(1+\epsilon^2)^{m+1/2}.
\label{eq6}
\end{equation}
In the $\epsilon\to 0$ limit, the second term in Eq.\eqref{eq6} goes to zero as
the Taylor expansion of $(1+\epsilon^2)^{m+1/2}$ around 
$\epsilon=0$ contains only even powers of $\epsilon$. Hence,
\begin{equation}
\lim_{\epsilon\to 0}I_{2m-1}=(2m)!i\pi ~.
\label{eq7}
\end{equation}
Similarly, when $n$ is even $n=2m$, we get
$$I_{2m}=i\frac{d^{2m+1}}{d\epsilon^{2m+1}}\left[\epsilon^2(1+\epsilon^2)^m
\mathcal{I}_1(\epsilon)\right],$$
as no term containing $K$-factors survives after taking the $(2m+1)$-th
derivative. Using the relation
$$ \mathcal{I}_1(\epsilon) \,=\,  \int\limits_{-\pi/2}^{\pi/2}
\frac{\cos{\theta}}{\sin^2{\theta}+\epsilon^2}
\;d\theta=\frac{2}{\epsilon}\arctan{\left(\frac{1}{\epsilon}\right)},$$
this equation reduces to 
\begin{equation}
I_{2m}=2i\frac{d^{2m+1}}{d\epsilon^{2m+1}}\left [\epsilon(1+\epsilon^2)^m
\arctan{\left(\frac{1}{\epsilon}\right)}\right ].
\label{eq8}
\end{equation}
Now we can take the limit $\epsilon\to 0$ by taking into account that 
$\arctan{\left(1/\epsilon\right)}$ and its
derivatives are not singular in this limit. Exploiting the Leibniz
rule to the Eq.(\ref{eq8}) one finds 
\begin{eqnarray}
\lim_{\epsilon\to 0}I_{2m} \,=\, 2i\lim_{\epsilon\to0}
\arctan{\left(\frac{1}{\epsilon}\right)}\frac{d^{2m+1}}{d\epsilon^{2m+1}}
[\epsilon(1+\epsilon^2)^m] \,=\,  i\pi(2m+1)! ~.~~
\label{eq9}
\end{eqnarray}
All other terms vanish because for any $j\ge 1$ either
$$\lim_{\epsilon\to 0}\frac{d^{2j}}{d\epsilon^{2j}}
[\epsilon(1+\epsilon^2)^m]=0 ~,$$
since binomial expansion of $\epsilon(1+\epsilon^2)^m$
contains only odd powers of $\epsilon$, or
$$\lim_{\epsilon\to 0}\frac{d^{2j}}{d\epsilon^{2j}}
\arctan{\left(\frac{1}{\epsilon}\right)}=-\lim_{\epsilon\to
0}\frac{d^{2j-1}}{d\epsilon^{2j-1}}\frac{1}{1+\epsilon^2}=0 ~.$$
Equations (\ref{eq7}) and (\ref{eq9}) can be unified in the final result
\begin{equation}
\lim_{\epsilon\to 0}I_n \,=\, i\pi (n+1)! ~.
\label{eq10}
\end{equation}

\section*{acknowledgments}

Authors are greatly indebted to Hendrik van Hees for useful discussions and to 
Igor Khavkine for his suggestion regarding the straightforward calculation of 
integrals for odd values of $n$ by using the Eq.\eqref{ganzogadtsarmoebuli}. 
Stimulating discussions with Gia Dvali and Cesar Gomez about the UV 
self-completeness of gravity are also kindly acknowledged. M.~M. is indebted 
to Marcus Bleicher and Piero Nicolini for their hospitality at the 
{\it Frankfurt Institute for Advanced Studies}. This research was supported in 
part by the National Research Fund, Luxembourg - AFR PhD 6758696 (A.~D.); by 
the Shota Rustaveli National Science Foundation under contract number 31/89 
and the DAAD research fellowship for university teachers and researchers 
(M.~M.); by the Ministry of Education and Science of the Russian Federation,
 Russian Federation President Grant for support of scientific schools 
NSh-2479.2014.2 and RFBR grant 13-02-00418-a (Z.~S.).

\end{document}